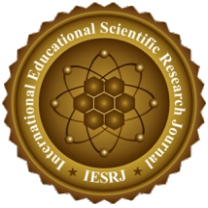

# ASTRONOMY LIKE THE FIRST CONTACT WITH SCIENCES

\* Vinicius de Abreu OLIVEIRA

Professor, Universidade Federal do Pampa, Caçapava do Sul - RS, Brazil – 96510-000. ( \*Corresponding Author)

## ABSTRACT

The study of the Universe and all its contents, the physical phenomena and the evolution of the sky objects has always attracted the Humanity. The Astronomy begins to try to explain these things, others sciences too. Therefore, we think a good first contact with the science world can be using a mobile inflatable planetarium and a portable telescope to show the amazing things about Astronomy. The first one allows us to show, in minutes, all the movements of the stars during a full night, explaining about observational Astronomy and so on. In addition to that, using a telescope we can show in real these movements and some interesting details about planets and constellations. The feedback from the students and teachers after the planetarium sections was important to us to evaluate the real necessity of works like that. The conclusion was directly: the whole audience loved the experience and almost all pretend go back in the next time.

## 1. Introduction

The study of Astronomy begins with the Humanity, after all, the curiosity and the search of knowledge has always been the motor for the development of the mankind. However, before the Astronomy is a formal science, it presented its bases either in mystical, cultural and religious reasons (Sagan, 1994).

Nowadays, more than 400 years after Galileu, considered the first professional astronomer, there is a great quantity of equipments to see the night sky, from lunettes to telescopes plugged on computers (Oliveira, 2014). Although the interest about Astronomy is always great, for the general sciences it is not true. Moreover, this interesting not necessary become a formal knowledge about the theme, re-feeding the mystical and cultural reasons about.

The National Curriculum Parameters in Brazil (Brazil, 1997, 1999) indicate guidelines to teach sciences, including Astronomy. However, the teacher's formation in sciences is poor because many factors (Saraiva & Kepler, 2009), which makes difficult the full implementation of the guidelines suggested.

Then, this work aims to help the teachers to attract young students to the amazing world of sciences using the Astronomy like the gateway to it.

## 2. Materials and Methods:

We divided the project in four steps: interview about the previous knowledge; a planetarium section; a new interview about what the students learned with; and, if possible, an observational night using a telescope of 150 mm.

The steps one and three are complementaries, using an Astronomy Questionnaire (AQ) developed by us to evaluate the basic knowledge about Astronomy and general sciences. The AQ was separated into age groups, respecting the knowledge expected in each one. However, the general context was always a comparison above the expectation with the planetarium and the real feeling about after the section.

The step two, in its turn, we used mobile inflatable planetarium (Figure 01) which enable us to take several schools during three weeks, with 20 students and one teacher, from small ages, or 15 adults.

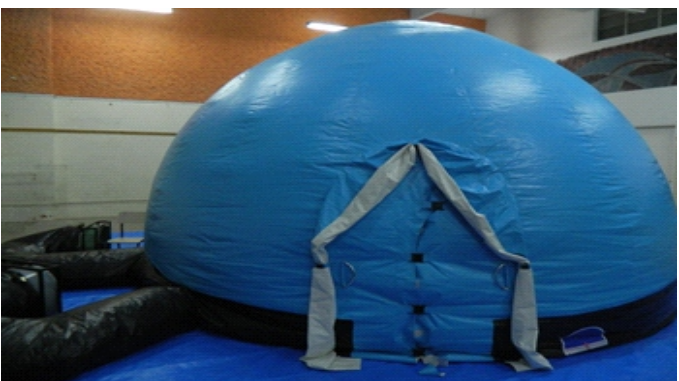

**Figure 01 – Our mobile inflatable planetarium, mounted in one school visited by us in 2015 May, Caçapava do Sul – RS, Brazil.**

Of course, the step four needs a good and dark night, without clouds and artificial lights. We managed to do only one, unfortunately.

## 3. Results:

Almost 1,800 people visited the planetarium during three weeks in 2015 May (Table 01), but the interview was made only with the students. Quantitatively were 21 schools, with students at all ages, from two cities, Caçapava do Sul and Lavras do Sul, both in the South of Brazil.

**Table 01 – Number of participants of the planetarium sections. The fourth column is the number of political and/or community authorities presents.**

| Students | Teachers | Community Resident | Others | Total |
|---|---|---|---|---|
| 1,596 | 115 | 70 | 15 | 1,796 |

On the whole, the students of the Brazil's equivalent to High School think that the planetarium help them to understand and learn the concepts about stars, constellations, planets and relative positions in the sky. In addition to understand the greatness of the Universe.

On the other hand, the students from the basic levels find the planetarium something between a TV show and magic. However, in general terms, they learn a little more about sciences and say they want to go back more often.

The observational night was terrific, we observed Venus and Jupiter, the main planets visible on that date. We pointed the telescope to Orion, The Warrior Constellation, and its nebula.

## 4. Discussion:

As a result of our AQ, it is possible to see a great excitement and anticipation about the planetarium, about the things they (the students) can see in there. Comparing the AQ before and the after, we noted that almost all the students had a scientific curiosity and a desire to understand more and more about the stars and the Universe.

It is noteworthy that some students had already showed the presentations in years before. They said that it is amazing and each time they learn something new. We think they probably will go into the next section too.

The observational night was perfect to show in locus how the sky objects move during the night, how observational astronomy works and etc. Using an informal and oral questionnaire we verified what was the most interesting thing during that night. The first one was the phases of the Planet Venus, because no one knew that Venus has phases like the Moon. The second was to see moons around Jupiter, a good manner to show that Earth is not in the center of the Universe.

## 5. Conclusions:

In order to help the teachers in the mission to teach about science and, at the same time, take the attention of the students, actions like this work is very useful. Astronomy is very visual and close to mankind, everyone looked at the stars and wondered. Because that, we think the use of a mobile inflatable planetarium and mobile telescopes are a good way to start. Especially for the students that have great difficulties to visit a fixed planetarium and big telescopes, like these.

**Acknowledgments:**
I would like to thank all the volunteers that help the execution of the work, in various levels. A special thank to Professor Guilherme Marranghello from UNIPAMPA, Bagé – RS, Brazil, because his help in logistic and with the planetarium with the project "Astronomia para todos".